\documentclass[epj]{svjour}
\usepackage{graphics}
\input babarsym
\def\allepsu{0.326}
\def\allepsmx{0.770} 

\def\allbrbrVal{2.06}
\def\allbrbrEcstat{0.25}
\def\allbrbrEsyst{0.23}
\def\allbrbrEthHi{0.36}

\def\allbr{ 2.24}
\def\allbrE{ 0.27}
\def\allbrEsyst{ 0.26}
\def\allbrEthHi{ 0.39}

\def\allvub{ 4.62}
\def\allvubE{ 0.28}
\def\allvubEsyst{ 0.27}
\def\allvubEthHi{ 0.40}
\def\allvubEtheo{ 0.26}

\def\cp {\ensuremath{C\kern-0.2em P}}

\newcommand {\breco}{\ensuremath{B_{reco}}}

\newcommand {\lbar}{\ensuremath{\overline{\Lambda}}}
\newcommand {\lone}{\ensuremath{\lambda_1}}
\newcommand {\ltwo}{\ensuremath{\lambda_2}}
\newcommand {\vcb}{\ensuremath{|V_{cb}|}}
\newcommand {\vub}{\ensuremath{|V_{ub}|}}

\newcommand {\Bxlnu}{\ensuremath{\Bb \to X \ell \bar{\nu}}}

\newcommand {\Bxclnu}{\ensuremath{\Bb \to X_c \ell \bar{\nu}}}

\newcommand {\Bxulnu}{\ensuremath{\Bb \to X_u \ell \bar{\nu}}}

\newcommand {\bsg}{\ensuremath{\b \to s\gamma}}

\newcommand {\mx}{\ensuremath{m_{X}}}

\newcommand {\meanmxx}{\ensuremath{\langle m_X^2 \rangle }}
\newcommand {\meanmx}{\ensuremath{\langle m_{X} \rangle }}

\newcommand {\rusl}{\ensuremath{R_{u}}}
\newcommand {\mmiss}{\ensuremath{m_{miss}^2}}

\newcommand {\bmeson}{$B$ meson}

\def\eg                 {{\it e.g.,}}
\def\etal               {{\it et~al.,}}

\begin{document}
\title{Inclusive Semileptonic  \mathversion{bold}$B$ Decays at \babar }
\author{Urs Langenegger (for the \babar\ Collaboration)
\thanks{Work supported in part by Department of Energy contract
  DE-AC03-76SF00515.
}%
}                     

\institute{Stanford Linear Accelerator Center, 
  Stanford University, Stanford, CA 94309, USA
}

%
\date{ } 
%
\abstract{ We report updates on two analyses of inclusive semileptonic
  $B$ decays  based on  a dataset of  89 million \BB\  events recorded
  with  the \babar\  detector at  the \FourS\  resonance.   Events are
  selected  by fully  reconstructing  the decay  of  one \bmeson\  and
  identifying a  charged lepton  from the decay  of the  other $\Bbar$
  meson.   In the  first analysis,  the measurement  of the  first and
  second moment  of the hadronic mass  distribution in Cabibbo-favored
  \Bxclnu\ decays allows for  the determination of the nonperturbative
  parameters \lbar\ and \lone\  of Heavy Quark Effective Theory (HQET)
  and \vcb.  In the second analysis, the hadronic mass distribution is
  used  to  measure  the  inclusive charmless  semileptonic  branching
  fraction and to determine \vub.
} 
\maketitle
%

\section{Introduction}
\label{intro}

The principal motivation for flavor physics is a comprehensive test of
the Standard  Model description  of \cp\ violation.   Semileptonic $B$
decays allow for the determination  of \vcb\ and \vub, two elements of
the  Cabibbo-Kobayashi-Maskawa  (CKM)  quark  mixing matrix.   In  the
unitarity  triangle,  constraints derived  from  kaon  decays and  the
overall normalization  depend on \vcb, while the  uncertainty in \vub\ 
dominates  the error  of the  length of  the side  opposite  the angle
$\beta$.  As this angle can be measured very cleanly in time-dependent
\cp\ asymmetries, the errors of  \vub\ must be model independent, well
understood,  and  small before  any  discrepancies  between sides  and
angles could be  interpreted as new physics.  Currently,  the error in
\vub\  is  dominated by  theoretical  uncertainties  in inclusive  $B$
decays and the absence of model independent formfactor calculations in
exclusive $B$ decays~\cite{Ligeti:2003hp}.

The   Cabibbo-favored  decays   \Bxclnu\  allow   for  high-statistics
measurements  of  HQET  parameters   and  quantitative  tests  of  the
consistency of  the underlying  theory.  The large  branching fraction
allows for clean experimental measurements with high purity event tags and
small systematic errors.  The  main difficulty in the determination of
\vub\ is  the large background from \Bxclnu\  decays, overlapping over
most of  the phase  space.  Selection cuts  reduce this  background by
restricting the phase  space, but lead to problems  in the theoretical
description.

The  measurements~\cite{Aubert:2003dr} presented here  are based  on a
sample of 89  million \BB\ pairs collected near  the \FourS\ resonance
by    the   \babar\   detector~\cite{Aubert:2001tu}.     The   boosted
center-of-mass system (CMS) at \babar\  leads to a limited coverage of
about $85\%$ of the solid angle  in the CMS.  The very high luminosity
opens alternative  methods in the precise study  of (semileptonic) $B$
decays.   Both analyses presented  here use  $\FourS \to  \BB$ events,
where  one $B$ meson  decays hadronically  and is  fully reconstructed
(\breco\ candidate)  and the semileptonic  decay of the  recoiling $B$
meson  is identified by  the presence  of an  electron or  muon.  This
approach  results in  a low  overall event  selection  efficiency, but
allows for  the determination of  the momentum, charge, and  flavor of
the $B$ mesons. To reconstruct  a large sample of $B$ mesons, hadronic
decays $\breco\to\Db  Y^{\pm}, \Db^* Y^{\pm}$ are  selected, where the
hadronic system $Y$  consists of $n_1\pi^{\pm}\, n_2K^{\pm}\, n_3\KS\,
n_4\piz$, with $n_1 + n_2 \leq 5$, $n_3 \leq 2$, and $n_4 \leq 2$. The
kinematic  consistency of  $B_{reco}$ candidates  is checked  with the
beam energy-substituted mass $\mes = \sqrt{s/4 - \vec{p}^{\,2}_B}$ and
the energy difference $\Delta E  = E_B - \sqrt{s}/2$, where $\sqrt{s}$
is  the  total energy  and  $(E_B,  \vec{p}_B)$  denotes the  momentum
four-vector of the $B_{reco}$ candidate in the CMS.

\section{Cabibbo-favored Decays \Bxclnu}
\label{sec:b2clnu}

Inclusive semileptonic  $B$ decays are  calculated in the  Heavy Quark
Expansion (HQE),  an Operator  Product Expansion using  HQET, allowing
for   the  computation   of,   \eg\  the   total  semileptonic   width
$\Gamma_{sl}$   in   terms  of   \vcb\   and   a   double  series   in
$\Lambda_{QCD}/m_b$ and $\alpha_s(m_b)$.  Higher order corrections are
parametrized  in  terms  of  expectation  values  of  hadronic  matrix
elements.   Other  observables,  \eg\  moments of  the  hadronic  mass
(squared) and lepton energy distributions, can be expressed in similar
expansions   with  different   dependences   on  the   nonperturbative
parameters.    An  overall   fit  to   these  moments   together  with
$\Gamma_{sl}$  thus provides  a  consistency check  of  the theory  by
comparing the  predicted and measured  moments and a  determination of
\vcb.   The  determination  of  the  $b$  quark  mass  $m_b$  and  the
parameters  $\lone$, $\ltwo$, $\rho_1$,  etc.  is  one of  the central
topics of semileptonic $B$ physics.  Special emphasis is placed on the
reduction of theoretical input and error estimates and to rely on data
from various processes.


This  analysis reports  an update  of the  measurement of  the moments
\meanmx\ and \meanmxx\ of the mass distribution of the hadronic system
$X$ in a semileptonic $B$ decay.

We select  events by requiring  a \breco\ candidate and  an identified
lepton  with momentum  in the  CMS $p^*  > 900  \mevc$, with  a charge
consistent  for a  primary  $B$  decay.  We  require  that the  charge
imbalance of the event is not larger than one and obtain a data sample
of about 7100 events.

We combine all  remaining charged tracks and neutral  showers into the
hadronic system  $X$.  A neutrino candidate is  reconstructed from the
missing  four-momentum $p_{miss}  =  p_{\FourS} -  p_X -  p_{\breco}$,
where all  momenta are  measured in the  laboratory frame.   We impose
consistency of  the measured  $p_{miss}$ with the  neutrino hypothesis
with the  requirements $E_{miss}  > 0.5\gev$, $|\vec{p}_{miss}|  > 0.5
\gev$,  and  $|E_{miss}  -   |\vec{p}_{miss}||  <  0.5\gev$.   A  $2C$
kinematic fit---imposing  four-momentum conservation, the equality of
the masses  of the two $B$  mesons and forcing  $p_\nu = 0$---improves
the resolution of the \mx\ measurement  to a width of about $350\mev$. 
Monte Carlo  simulated event  samples are used  to calibrate  the mass
scale, determine efficiencies, and estimate backgrounds.

\begin{figure}
 \resizebox{0.5\textwidth}{!}{
   \begin{picture}(40.,23.)
    \put(0.0, -2.0){\mbox{\includegraphics{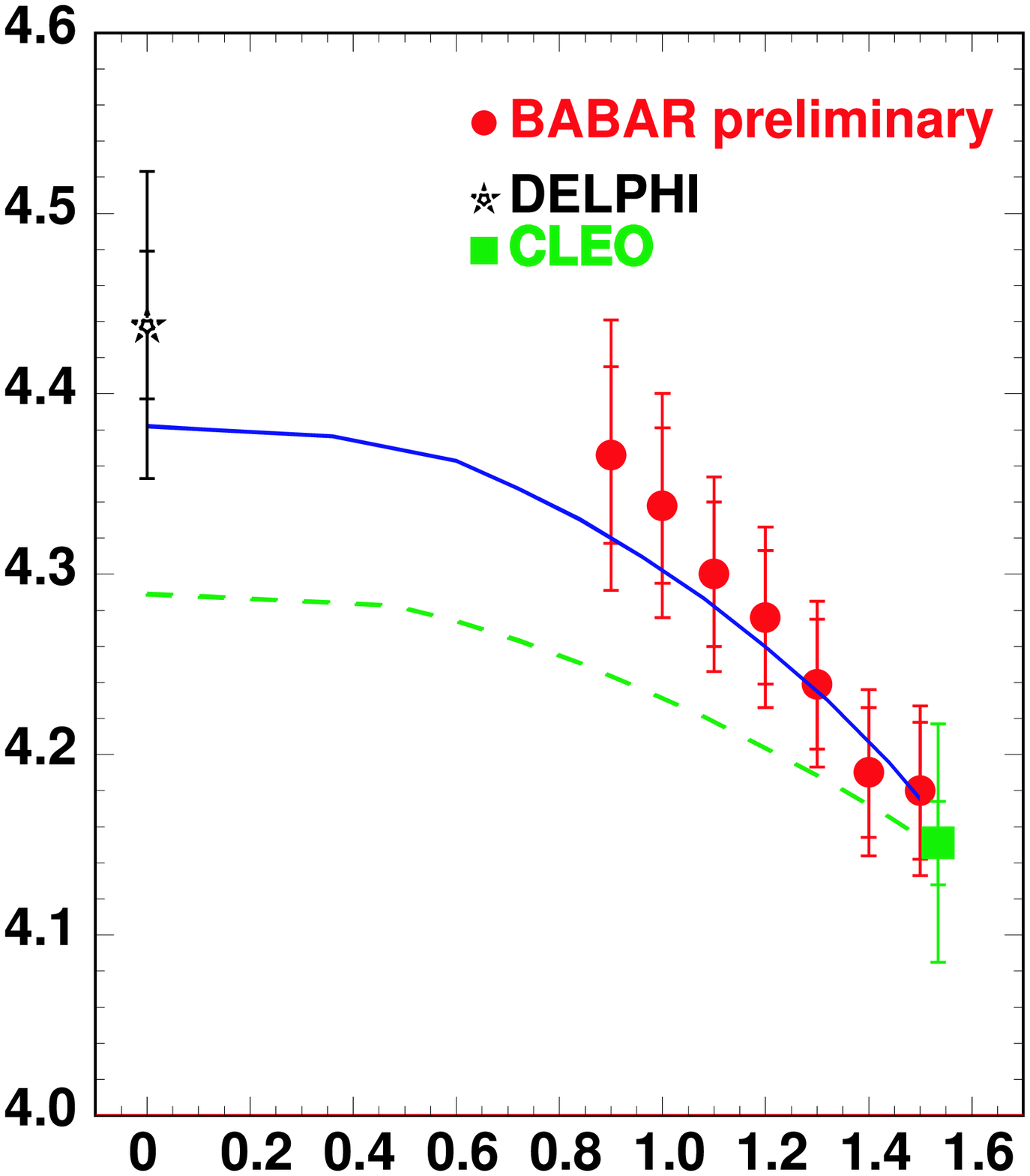}}}
    \put(19.0,1.4){\resizebox{22.5cm}{!}{\includegraphics{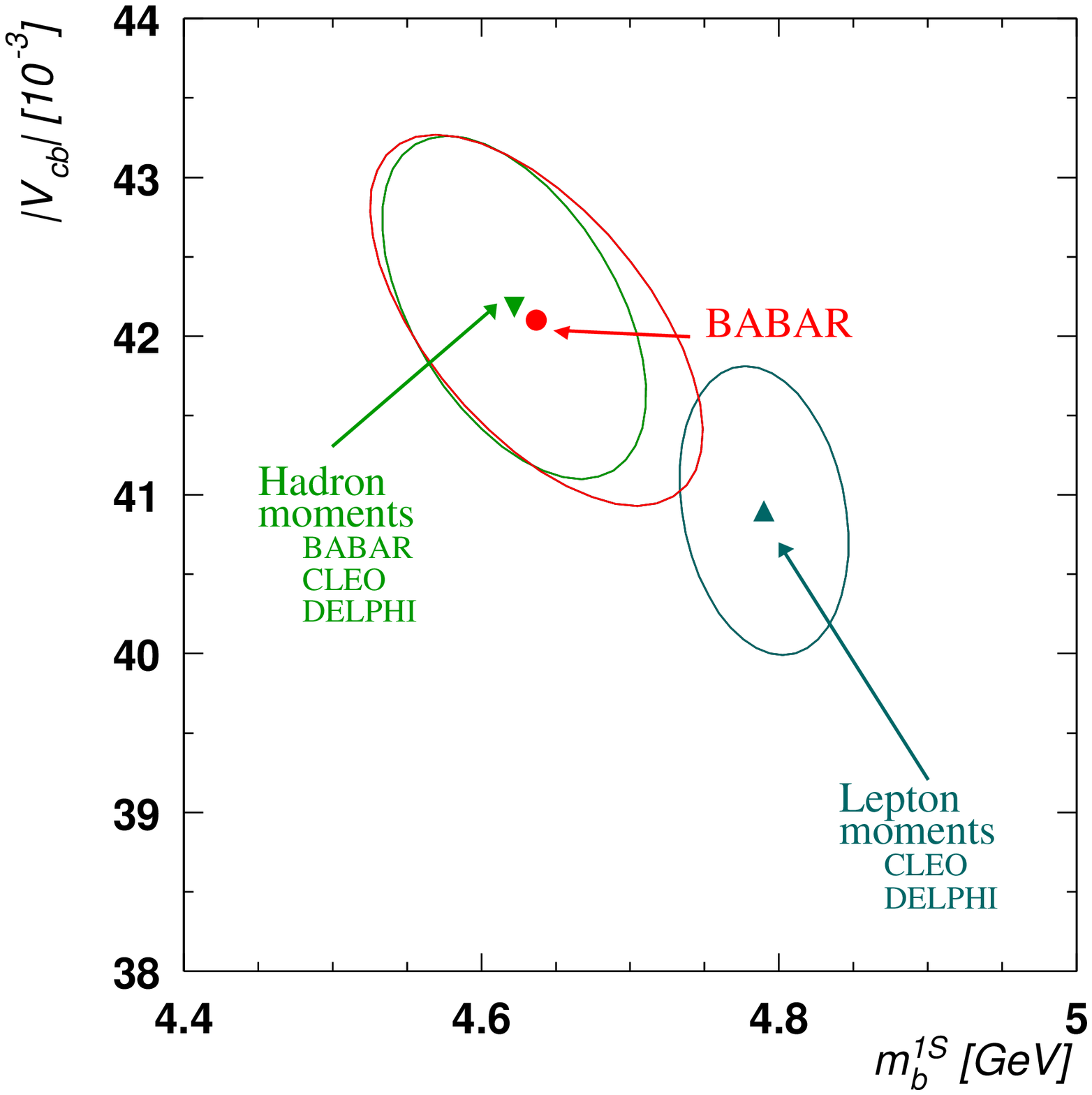}}}
    \put(14.8,1.8){{\Huge $p^*_{min} [\gev]$}}
    \put(0.5,17.0){\rotatebox{90}{\Huge $\langle m_X^2\rangle [\gev^2]$}}
    \put(4.6, 18.2){\resizebox{4.5cm}{!}{\includegraphics{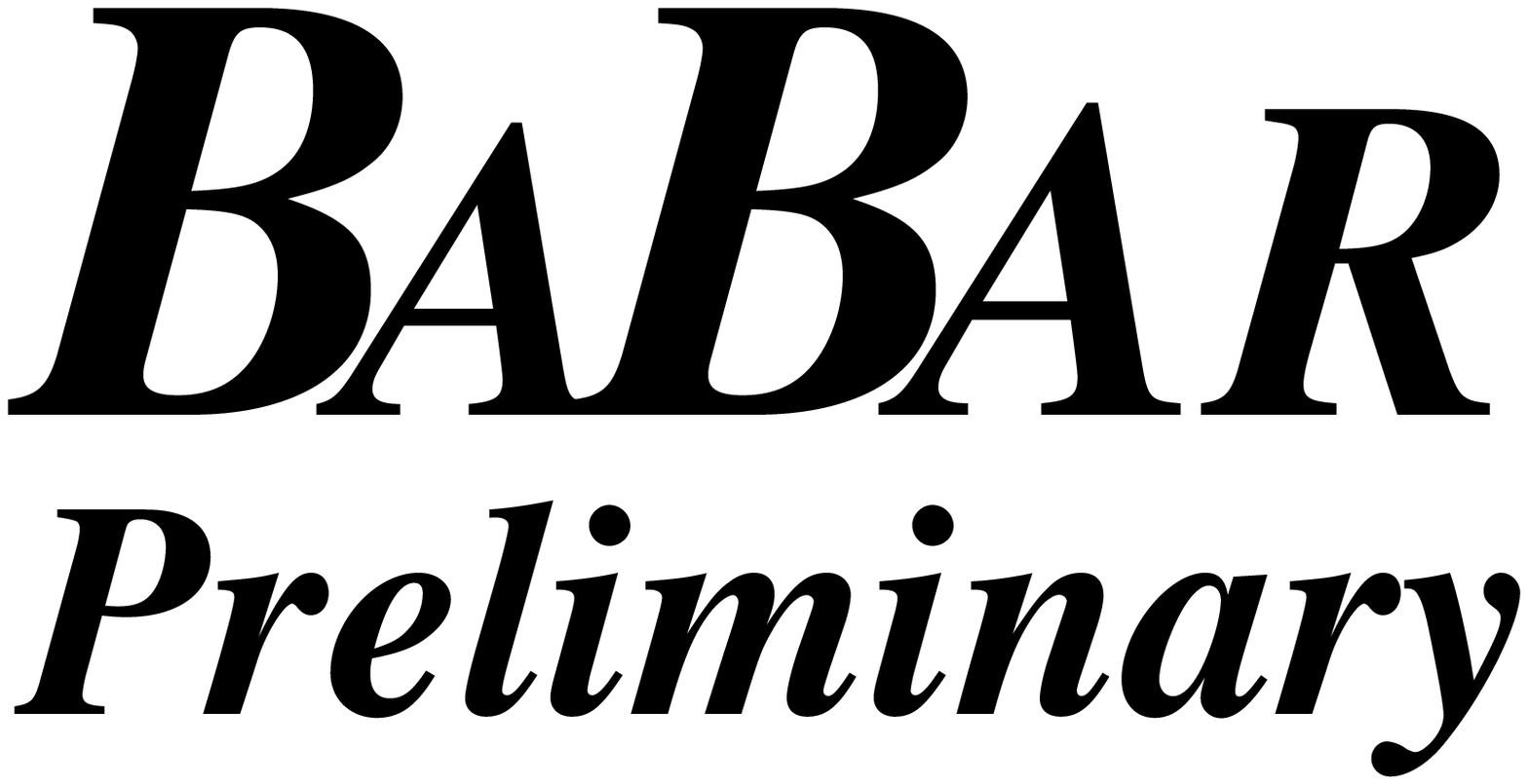}}}
    \put(34.0, 18.2){\resizebox{4.5cm}{!}{\includegraphics{babarPrel.eps}}}
    \put(4.0, 20.0){{\Huge a)}}
    \put(23.0,20.0){{\Huge b)}}
   \end{picture}
 }
 \caption{(a) Measured moments \meanmx\ for different lepton threshold
   momenta  $p^*_{min}$.    The  errors  of   the  individual  \babar\ 
   measurements   are   highly   correlated.   For   comparison,   the
   measurements       by       the      DELPHI~\cite{delphi}       and
   CLEO~\cite{Cronin-Hennessy:2001fk}  collaborations are also  shown. 
   The solid curve  is a fit to the \babar\ data;  the dashed curve is
   the        prediction        based        on        the        CLEO
   results~\cite{Cronin-Hennessy:2001fk,Chen:2001fj}.  (b) Constraints
   on the  $b$ quark mass  and \vcb\ from  the combined fit  to hadron
   moments and lepton moments, respectively.
  \label{fig:1}       
}
\end{figure}

The resulting  moments of  the hadronic mass-squared  distribution are
shown as  a function of  the threshold lepton momentum  $p^*_{min}$ in
Fig.~\ref{fig:1}a.  A  substantial rise  of the moments  towards lower
momentum is  visible, due to  the enhanced contributions  of high-mass
charm states (phase-space suppressed  at higher $p^*_{min}$). The main
contributions  to  the systematic  error  are  the  simulation of  the
detector  response  and  residual  backgrounds.   The  uncertainty
from the modeling  of the  $X_c$ state  is negligible compared to the
other systematic errors.


Accounting  for  all correlations  between  the  moments of  different
$p^*_{min}$,  we determine  $\lbar^{\overline{MS}}  = 0.53\pm0.09\gev$
and    $\lone^{\overline{MS}}    =    -0.36\pm0.09\gev^2$    in    the
$\overline{MS}$  scheme~\cite{Falk:1997jq}.  The  errors given  do not
include   uncertainties  due  to   terms  ${\cal   O}(1/m_B^3)$.   For
comparison,  we  also show  in  Fig.~\ref{fig:1}a  the  result of  the
hadronic  mass measurement  of DELPHI~\cite{delphi},  fully consistent
with  our result.  The  CLEO result~\cite{Cronin-Hennessy:2001fk}
of the  first hadronic  mass moment at  $p^*_{min} = 1.5\gev$  is also
consistent  with our  measurement, but  in combination  with  the mean
photon energy from $b\to s\gamma$~\cite{Chen:2001fj} shows a different
$p^*_{min}$   dependence  (see   Ref.~\cite{Bigi:2003zg}   for  recent
developments).

The  calculations  of Ref.~\cite{Bauer:2002sh}  are  used  to fit  all
hadronic  moments from  \babar\ in  the  $1S$ scheme,  as this  scheme
exhibits better convergence of  the series than other alternatives. We
find $m_b^{1S}  = 4.638 \pm 0.094_{exp} \pm  0.062_{dim\oplus BLM} \pm
0.065_{1/m_B^3}   \gev$   and  $\lone   =   -0.26\pm  0.06_{exp}   \pm
0.04_{dim\oplus BLM} \pm 0.04_{1/m_B^3}  \gev^2$. In this fit, we take
into  acount  all correlations  between  the  experimental results,  a
significant   improvement   with    respect   to   the   approach   of
Ref.~\cite{Bauer:2002sh}.   The  fit  also utilizes  the  semileptonic
width  $\Gamma_{sl}  =  (4.37\pm 0.18)\times10^{-11}\mev$  (determined
from \babar\  data) and determines  $\vcb = (42.10 \pm  1.04_{exp} \pm
0.52_{dim\oplus BLM} \pm 0.50_{1/m_B^3})\times 10^{-3}$.

We test  the consistency  of the HQE  by combining the  measurement of
\babar\  with the  four lepton  energy  moments measured  by the  CLEO
collaboration~\cite{Briere:2002hw}   and  the  hadronic   mass  moment
measurement    of   the   DELPHI    collaboration~\cite{delphi}.    In
Fig.~\ref{fig:1}b,  the fit  results are  shown separately  for hadron
mass and  lepton energy moments.   The $\Delta\chi^2 = 1$  contours of
hadronic mass and  lepton energy moments do not  overlap.  The largest
errors in these measurements are due to the unknown higher order terms
of  order $1/m_B^3$.   

\unitlength1.0cm 


\section{Cabibbo-suppressed Decays \Bxulnu}
\label{sec:vub}

In  the measurement  of  \Bxulnu\ decays,  the  large background  from
\Bxclnu\  decays  is traditionally  reduced  by  measuring the  lepton
spectrum at the ``endpoint'', beyond the kinematic cutoff for \Bxclnu\ 
decays.  A  disadvantage of this approach  is that only  about 10\% of
all  charmless  semileptonic  decays  are  measured.   This  leads  to
significant  extrapolation uncertainties,  which can  be  reduced with
information  on the movement  of the  $b$ quark  inside the  $B$ meson
obtained from the photon energy spectrum in \bsg\ decays.

Here we use the invariant mass \mx\ of the hadronic system to separate
\Bxulnu\       decays      from       the       dominant      \Bxclnu\ 
background~\cite{Barger:tz}. This method offers a substantially larger
acceptance than  the endpoint measurement.  As in  the first analysis,
the hadronic  system $X$  in the decay  \Bxlnu\ is  reconstructed from
charged  tracks   and  energy  depositions  in   the  calorimeter  not
associated with the \breco\ candidate or the identified lepton.
We require  exactly one  charged lepton with  $p^* > 1  \gevc$, charge
conservation ($Q_{X}  + Q_\ell + Q_{\breco}  = 0$), and  $\mmiss < 0.5
\gev^2$.   
We reduce  the $\Bzb\to\Dstarp\ell^-\overline{\nu}$ background  with a
partial  reconstruction   of  the   decay  (the  $\pi^+_s$   from  the
$\Dstarp\to \Dz\pi_s^+$  decay and the lepton).   Furthermore, we veto
events with charged or neutral kaons in the recoil \Bb.


In order  to reduce experimental  systematic errors, we  determine the
ratio of branching fractions \rusl\ from $N_u$, the observed number of
$\Bxulnu$   candidates   with    $\mx<1.55$\gevcc,   and   $N_{sl}   =
29982\pm233$, the number of events with at least one charged lepton:
\begin{displaymath}
\rusl=
\frac{\BR(\Bxulnu)}{\BR(\Bxlnu)}=
\frac{N_u/(\varepsilon_{sel}^u \varepsilon_{\mx}^u)}{N_{sl}} 
\times \frac{\varepsilon_l^{sl} \varepsilon_{reco}^{sl} } {\varepsilon_l^u \varepsilon_{reco}^u }.
\label{eq:vubExtr}
\end{displaymath}
Here $\varepsilon^u_{sel}  = \allepsu\pm0.6_{stat}$ is  the efficiency
for  selecting  \Bxulnu\ decays  once  a  \Bxlnu\  candidate has  been
identified,  $\varepsilon^u_{\mx}  =  \allepsmx\pm0.9_{stat}$  is  the
fraction    of    signal    events    with   $m_X    <    1.55\gevcc$,
$\varepsilon_l^{sl}/\varepsilon_l^u  =  0.887\pm0.008_{stat}$ corrects
for the  difference in the efficiency  of the lepton  momentum cut for
\Bxlnu\           and          \Bxulnu\           decays,          and
$\varepsilon_{reco}^{sl}/\varepsilon_{reco}^u =  1.00 \pm 0.03_{stat}$
accounts  for  a  possible  efficiency difference  in  the  $B_{reco}$
reconstruction in  events with \Bxlnu\ and \Bxulnu\  decays. 

\begin{figure}
\resizebox{0.5\textwidth}{!}{%
  \includegraphics{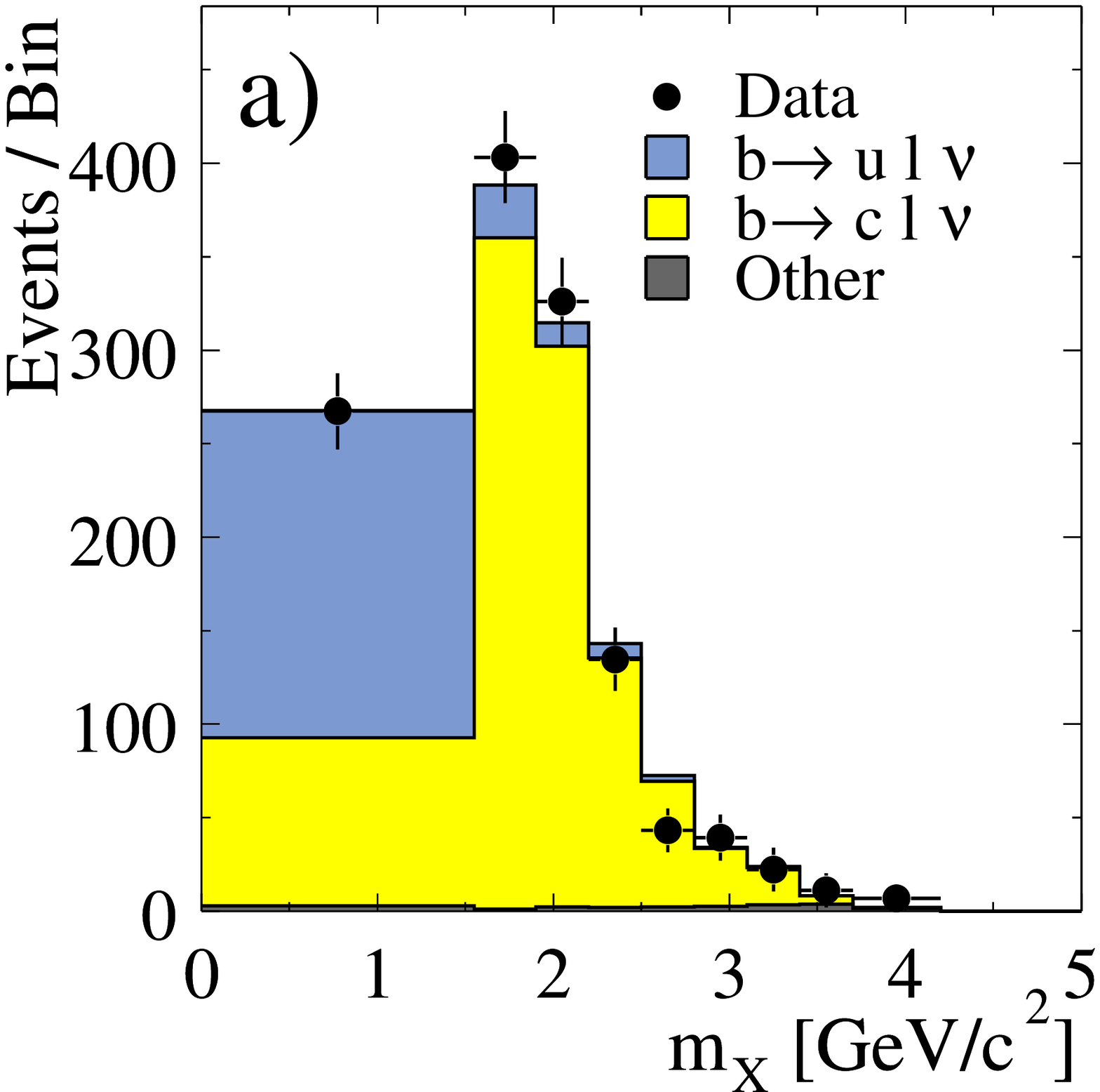}  \includegraphics{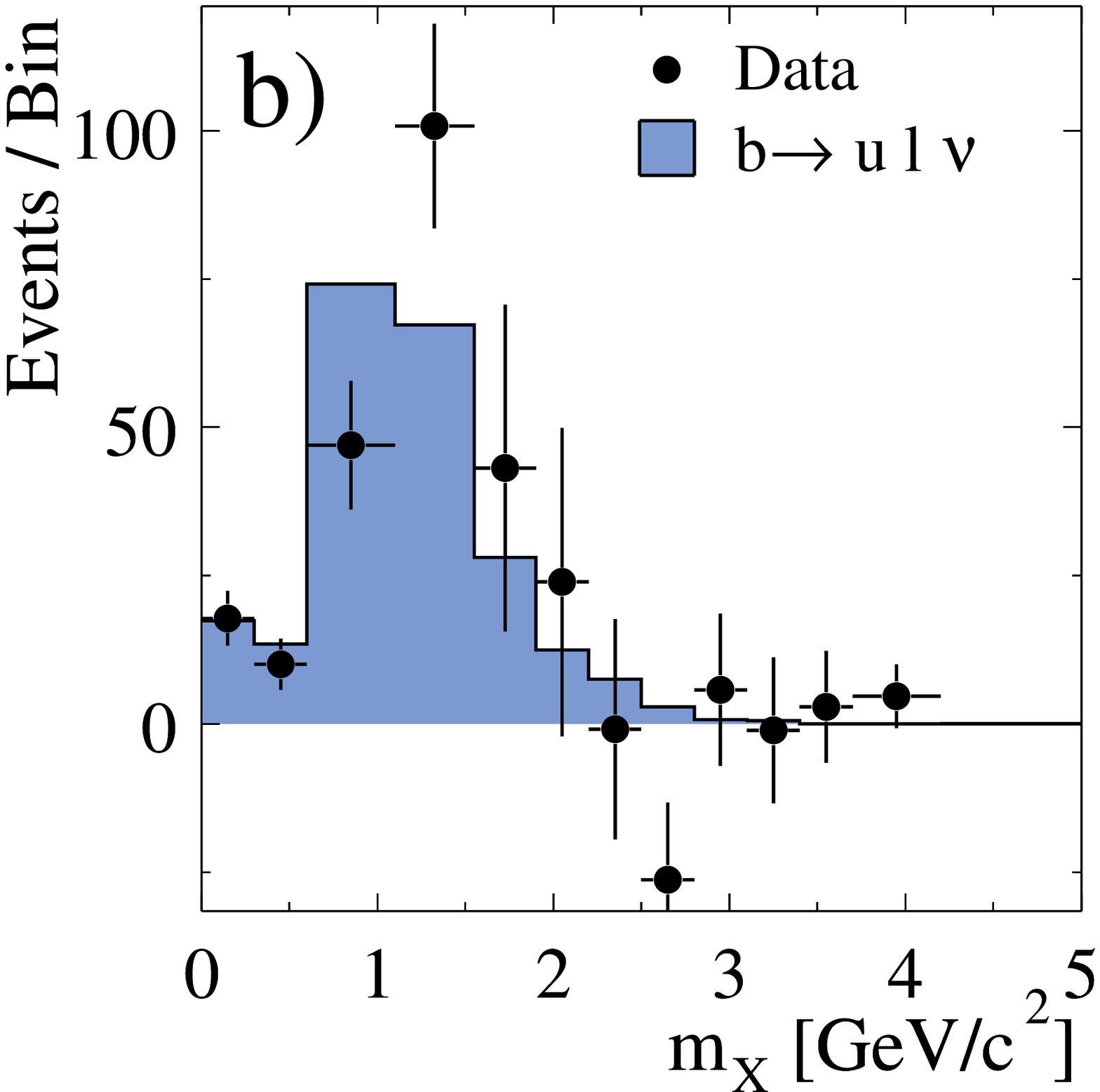}
}
\caption{The $\mx$ distribution for \Bxlnu\ candidates: a)
  data (points) and fit components, and b) data and signal MC after
  subtraction of the $b\to c\ell\nu$ and the ``other'' backgrounds.
  \label{fig:2}       
}
\end{figure}

We extract  $N_u$ from the $\mx$ distribution  by a fit to  the sum of
three contributions:  signal, background  $N_{c}$ from \Bxclnu,  and a
background  of $<1\%$ from  other sources.   
Fig.~\ref{fig:2}a  shows the fitted  $\mx$ distribution.   To minimize
the model dependence, the first bin is extended to $\mx < 1.55\gevcc$.
We find
$175\pm 21$ signal events and $90\pm5$ background events in the region
$\mx < 1.55\gev$.

The dominant detector systematic  errors are due to the uncertainty in
photon  detection and   combinatorial background  subtraction.  
We assess the theoretical uncertainties by varying the nonperturbative
parameters  in Ref.~\cite{DeFazio:1999sv}  within  their errors,
$\lbar  = 0.48  \pm  0.12  \gev$ and  $\lone  = -0.30\pm0.11  \gev^2$,
obtained  from the  results  in Ref.~\cite{Cronin-Hennessy:2001fk}  by
removing  terms proportional  to $1/m_b^3$  and $\alpha_s^2$  from the
relation between the measured observables and \lbar\ and \lone.

In summary, we have
$\rusl = (\allbrbrVal \pm \allbrbrEcstat
\pm \allbrbrEsyst
\pm\allbrbrEthHi)\times 10^{-2}.
$
Combining the  ratio \rusl\  with the measured  inclusive semileptonic
branching fraction of Ref.~\cite{Aubert:2002uf}, we obtain
$\BR(\Bxulnu) = 
(\allbr
\pm\allbrE
\pm\allbrEsyst
\pm\allbrEthHi)\times10^{-3}$. 
With  the   average   $B$  lifetime   of Ref.~\cite{pdg2002} we obtain
$
\vub = (\allvub
\pm\allvubE
\pm\allvubEsyst
\pm\allvubEthHi     \pm\allvubEtheo)\times    10^{-3}$     based    on
Ref.~\cite{Uraltsev:1999rr}.   The  first  error is  statistical,  the
second systematic, the third  gives theoretical (signal efficiency and
the extrapolation of  \rusl\ to the full $\mx$  range), and the fourth
is the  uncertainty in  the extraction of  \Vub\ from the  total decay
rate. No error is assigned to the assumption of parton-hadron duality.
This result is consistent with previous inclusive measurements, but
has a smaller systematic error, primarily due to larger
acceptance and higher sample purity.  The results of exclusive
measurements tend to have a lower central value, but with
a slightly larger error.

\section{Outlook}
\label{sec:concl}

Both analyses presented here  will benefit from higher statistics.  By
measuring higher moments of the  hadronic mass, the lepton energy, and
possibly  other distributions,  the analysis  of \Bxclnu\  decays will
gain sensitivity to  higher order HQET parameters and  thus reduce the
theory  dependent error  in  the  determinations of  \vcb\  and \vub.  
Independent measurements  of $  b \to s$  transitions are  expected to
provide another means of constraining the theoretical uncertainties.

%
%
%

\end{document}